\documentstyle{article}
\begin{document}
\LARGE
\begin{center}
\vspace*{0.3in} \bf Creation of Kerr-de Sitter Black Hole in All Dimensions
\vspace*{0.6in} \normalsize \large

\rm Zhong Chao Wu

DAMTP, University of Cambridge,
Wilberforce Road,
Cambridge CB3 0WA, UK

and

Dept. of Physics,
Zhejiang University of Technology,
Hangzhou 310032, China

\vspace*{0.4in} \large \bf Abstract
\end{center}
\vspace*{.1in} \rm \normalsize \vspace*{0.1in}

We discuss quantum creation scenario of Kerr-de Sitter black hole in
all dimensions. We show that its relative creation probability is the
exponential to the entropy of the black hole, using a topological
argument. The action of the Euclidean regular instanton can be readily
calculated in the same way.

 \vspace*{0.3in}

PACS number(s): 04.70.Dy; 98.80.Hw; 98.80.Bp; 04.60.Kz


\vspace*{0.5in}

e-mail: zcwu@zjut.edu.cn,

\pagebreak

In gravitational physics, black holes and the universe are the two
most attractive objects. Therefore, a quantum creation of black
hole at the birth of the universe is one of the most interesting
problems in the field. Extensive work has been done for the
creation scenario for all kinds of black holes in a 4-dimensional
spacetime. Most work is done for the maximal black hole creation
[1], although the non-extreme sub-maximal black hole creation has
also been discussed [2][3]. The case of non-rotating black holes
in all dimensions can be simply treated, since its higher
dimensional metric is a quite straightforward generalisation of
the 4-dimensional one [2][4].

The Kerr solution was found about half century after the discovery
of general relativity [5]. Two decades later Myers and Perry were
able to get its higher dimensional generalisation in the flat
background [6]. Only very recently the metrics of $5$-dimensional
and higher dimensional Kerr black holes in the de Sitter
background were found [7][8].

The creation scenario of general 4-dimensional Kerr-Newman black holes in
the de Sitter, flat and Anti-de Sitter backgrounds has been tackled, it turns out that
the creation probability in the closed (open) background is the
exponential to (the negative of) the entropy of the created
hole [2]. The creation of 4-dimensional charged rotating extreme black holes from a
regular seed instanton was discussed in [9].

In the $D$-dimensional Kerr-de Sitter spacetime there are $N
\equiv [(D-1)/2]$ rotation parameters $a_i$ for orthogonal planes,
the rotations are associated with the azimuthal coordinates
$\phi_i$ with periods $2\pi$ and latitudinal coordinates $\mu_i$,
which lie in the interval $0 \le \mu_i \le 1$. One can set $n
\equiv [ D/2]$ and $\epsilon = n-N$. If $D$ is even, then
$\epsilon = 1$ , $n = N +1$ and there is an extra $\mu_{N}$, for
which $-1 \le \mu_{N+1} \le 1$. The sum of all the $\mu_i^2$
should be unity.

The Kerr-de Sitter metric with cosmological constant $\Lambda$ can be
cast in the  Boyer-Lindquist coordinate [8]
\[
ds^2 = -W (1- \lambda r^2)dt^2 + \frac{Udr^2}{V-2M} + \frac{2M}{U}
\left ( dt - \sum_{i =1}^N \frac{a_i\mu_i^2 d\phi_i}{1 + \lambda
a_i^2} \right )^2
\]
\[
+ \sum_{i =1}^n \frac{r^2 + a_i^2}{1 + \lambda a_i^2} d\mu^2_i + \sum_{i =1}^N \frac{r^2 + a_i^2}{1 + \lambda a_i^2} \mu^2_i (d\phi_i - \lambda a_id t)^2
\]
\begin{equation}
+ \frac{\lambda}{W(1- \lambda r^2)} \left ( \sum_{i =1}^n \frac{(r^2 + a^2_i)\mu_id\mu_i}{1 + \lambda a^2_i} \right )^2,
\end{equation}
where $M$ is the mass parameter, $\lambda \equiv  \frac{2 \Lambda}{(D-1)(D-2)}$ and
\begin{equation}
W = \sum_{i=1}^n \frac{\mu_i^2}{1 + \lambda a_i^2},
\end{equation}
\begin{equation}
U = r^{\epsilon}\sum_{i=1}^n\frac{\mu_i^2}{r^2 + a_i^2} \prod_{j=1}^N (r^2 + a_j^2)
\end{equation}
and
\begin{equation}
V = r^{-2+\epsilon}(1-\lambda r^2) \prod_{i =1}^N(r^2 + a_i^2)
\end{equation}

For the creation of the above  higher dimensional Kerr black hole in the de
Sitter background, one needs to find the seed instanton. The instanton can be obtained through the analytic
continuation $\tau = it$ from metric (1). The horizons are
located at the zeroes of $V-2M = 0$. The instanton is constructed by
some periodic identification of coordinate $\tau$ between two
horizons $r_l \; (l= 1,2)$, which are the two largest
zeroes of $V-2M=0$ and we set $r_2 > r_1$.

To obtain a regular instanton $M$, one needs to identify $\tau$ by a
period of $\beta_l = 2\pi/\kappa_l$ in order to avoid  the conical singularity
at horizon $r_l$, where $\kappa_l$ is its surface gravity [8]
\begin{equation}
\kappa_l = r_l (1-\lambda r_l^2) \left ( \sum_{i =1}^N \frac{1}{r_l^2 +
  a^2_l} + \frac{\epsilon}{2r_l^2} \right ) - \frac{1}{r_l}.
\end{equation}
To avoid the conical singularities for both the two horizons, their surface
gravities must take the same value $\kappa$. This condition can be met only
for a degenerate case of metric (1).

To avoid the irregularities caused by the differential rotation of two
horizons, the following condition should also be satisfied [10]
\begin{equation}
(\Omega^j_1 - \Omega^j_2) \frac{2\pi i}{\kappa} = 2m_j \pi,
\end{equation}
where $m_j$ are integers, and $\Omega^j_l$ are the angular velocities of the horizon $r_l$ [8]
\begin{equation}
\Omega^j_l = \frac{a_j (1 + \lambda a^2_j)}{r_l^2
  +a^2_j}.
\end{equation}
It turns out that the smooth compact
regular instanton is $S^{D-2}$ bundles over
$S^2$. In particular, there are infinitely many for each odd $D\ge
5$ [8]. It is noted that in order to get an Euclidean instanton, the
Lorentzian angular momentum parameters $a_j$ should take an imaginary
value. However, we shall use an alternative prescription below, that is $a_j$
remains to be real.

We are interested in the creation of the black hole from a more
general instanton, i.e. the constrained instanton, in which the
above restrictions (6)(7) are relaxed. Of course, the creaction of
a maximal black hole is the special case of the consideration
here.  We choose an arbitrary period $\beta$, the identification
we are using is $ \tau \longrightarrow \tau + \beta, \;
r\longrightarrow r, \mu_i \longrightarrow \mu_i, \phi_i
\longrightarrow\phi_i + \eta_i$, where $\eta_i$ are arbitrary
constants. The obtained  metric is complex. Since this
identification does not meet the regularity conditions for the
regular instanton, there is at least one conical singularity
associated with one of the horizons, and also other irregularities
associated with the differential rotations there.

The quantum transition should occur at one equator, say the joint sections $\tau
=0$ and $\tau = \beta/2$. The created Lorentzian spacetime metric can
be obtained via the reversal analytic continuation at the equator. Strictly
speaking, the seed instanton should be constructed by a south
``hemisphere'' with its time reversal, the north
``hemisphere''. However, this consideration will not affect our
calculation below.  The
relative creation probability of the black hole, at the $WKB$
level, is [11]
\begin{equation}
P \approx \exp - I,
\end{equation}
where $I$ is the Euclidean action of the instanton. It is written as
\begin{equation}
I = -\frac{1}{16 \pi}\int_M({}^DR - 2\Lambda) -
\frac{1}{8\pi}\int_{\partial M}{}^{D-1}K,
\end{equation}
where ${}^DR$ is the scalar curvature of the $D-$dimensional spacetime
and ${}^{D-1}K$ is the extrinsic curvature of its boundary. The second
term includes contributions from the discontinuities of the
extrinsic curvature and the conical singularity (as a degenerate form).

One could calculate the action directly, as for the Kerr-Anti-de
Sitter black hole case [12]. In the Kerr-Anti-de Sitter case one
has to perform the background subtraction to regularize the
expression. Here, we use an alternative method, that is a
topological argument, which is more transparent due to the
topological  origin of the entropy. The most convenient way is to
divide  $M$ into three parts: $M_1 \; (r_1 \le r \le r_1 +
\delta)$, $M_2 \; (r_2- \delta \le r \le r_2)$ and $M^\prime \;
(r_1 + \delta \le r \le r_2 - \delta)$, where $\delta$ is a
positive infinitesimal quantity. The total action is
\begin{equation}
I = I_1 + I_2 + \int_{M^\prime} (\pi^{pq}\dot{h}_{pq} - N H_0 -N_pH^p)
d^{D-1} xd \tau ,
\end{equation}
where  $I_i$ is the action for the submanifold $M_i$. The third term of
the action for $M^\prime$  has been cast into the canonical
form in the  $1 + (D-1)$ decomposition. Here, $\pi^{pq}$ is the
canonical momentum conjugate to the metric $h_{pq}$ of hypersurfaces $\tau =
consts.,$
$H_0$ and $H^p$ are the Einstein and momentum constraints, which
should be zero for the instanton. The dot denotes the time derivative, it must vanish due to the
$U(1)$ symmetry associated with the Killing coordinate
$\tau$. Therefore, the third term of the action must be zero.

To calculate the first or second term of the action, one can apply the
Gauss-Bonnet theorem to the 2-dimensional $(\tau, r)$ section of
$M_l$ under the condition that other coordinates are frozen,
\begin{equation}
\frac{1}{4\pi}\int_{\hat{M_l}}{}^2R + \frac{1}{2\pi}\int_{\partial
  \hat{M}_l}{}^1K + \frac{\sigma}{2\pi} = \chi (l),
\end{equation}
where $\hat{M}_l$ represents the section, ${}^2R$ and ${}^1K$ are the
scalar curvature of $\hat{M}_l$ and extrinsic curvature of $\partial
  \hat{M}_l$, respectively, $\sigma$ is the deficit angle of the
  conical singularity at the horizon or the apex, and $\chi (l)$ is the Euler
  characteristic of $\hat{M}_l$, which is $1$ here.

The first terms of both (9) for $M_l$ and (11) are zero
as $\delta$ approaches zero. At the horizon one has ${}^{D-1}K= {}^1K$, and the deficit
angle is included as a degenerate form in the second term of (9),
Comparing (9) and (11), one obtains
\begin{equation}
I_l = - \frac{1}{4} A_l,
\end{equation}
where $A_l$ is the horizon area. Its value is given [8]
\begin{equation}
A_l = \frac{2\pi^{(D-1)/2}}{\Gamma [(D-1)/2]} r_l^{\epsilon -1} \prod_{i
  =1}^{N} \frac{r_l^2 + a^2_i}{1 + \lambda a_i^2}.
\end{equation}

One has to be cautious for the decomposition in (10), we have
implicitly assumed that $M_1$ and $M_2$ are regular everywhere except the conical
singularities, this is only possible when $\phi_i - \Omega^i_l \tau$
are fixed at a fixed point at the horizon $r_l$. However, the angular
velocities at the two horizons are different in general. Therefore, there must be
discrepancies of
the coordinate angle $\phi_i$ and discontinuity of extrinsic curvature across $r = r_1 + \delta$ and $r =r_2 -
\delta$, for any identification parameters $\eta_i$. As $\delta$ approaches zero, all these irregularities will
be located at the horizons, and the configuration of the wave function at
the equator remains intact.  This part of
contribution should be cancelled by adding the Legendre terms, that is
the differential rotating angles of the two horizons times their corresponding angular momenta, as
we did earlier [2]. Using the canonical form, this problem has been
taken care of automatically. Since the third term of the action is
additive, it means that the Gibbons-Hawking boundary term has been
implicitly included. If one naively integrates the volume term of (9) for $M^\prime$, the
result would not be zero. One can
rephrase this phenomenon by saying that the angle is not the  right
representation for the metric of the equator in computing the creation probability. Instead, their conjugate variables, the angular
momenta are the right representation. This problem did not occur for
the earlier research with other regular instantons, since then the relevant
Legendre terms vanished anyway.

Therefore, the total action is one quarter of the sum of the two
horizon areas. Since we have shown
the constructed manifold satisfies the field equation everywhere
except at the horizons, for the given equator configuration the only degree
freedom left are $\beta$ and $\eta_i$. We have shown that the action is independent of
these parameters. This means that the action is stationary with
arbitrary variation with the restrictions at the equator.
Therefore it is qualified as a constrained instanton and formula (8)
is valid.

In gravitational thermodynamics [13] the Euclidean path integral
can be interpreted as the partition function. Our right
representation corresponds to the microcanonical ensemble. In this
ensemble the entropy $S$ can be obtained from the partition
function $Z$
\begin{equation}
S = \ln Z \approx -I,
\end{equation}
the approximation is at the $WKB$ level. Therefore, the relative
creation probability is the exponential to the entropy of the created
hole. This is the result we wish to show.

Our method of computing the action using canonical form can also be
applied to the Euclidean regular instanton [8], in which the two roots for
the horizons coincide. For the selected parameters and the right
angular velocity for the identification as required by (6), one can avoid all
irregularities, and then the action should be the negative of half
of the horizon area, as in the case of usual 4-dimensional Nariai instanton.

It is clear that the method used here is also applicable to all
stationary metric with at least two horizons, in particular to the
metric of the higher dimensional charged Kerr-de Sitter black
hole, which has been found for $5-$dimensional case [14] and is
expected to be found in all dimensions in the near future.

 \vspace*{0.3in} \rm

\bf Acknowledgement:

\vspace*{0.2in} \rm

I would like to thank S.W. Hawking for his hospitality and
G.W. Gibbons for useful discussions.

\vspace*{0.3in} \rm

\bf References:

\vspace*{0.2in} \rm

1. See, for example, S.F. Ross, in \it The Future of Theoretical Physics and Cosmology,
Celebrating Stephen Hawking's 60th Birthday, \rm eds. G.W. Gibbons,
E.P.S. Shellard and S.J. Rankin, Cambridge University Press
(2003) and references therein.

2.  Z.C. Wu, \it Int. J. Mod. Phys. \rm \bf D\rm\underline{6}, 199
(1997);  \it Phys. Lett. \rm \bf
B\rm \underline{445}, 274 (1999);  \it Gene. Relativ. Grav. \rm\underline{32}, 1823 (2000).

3.  R. Bousso and S.W. Hawking, \it Phys. Rev. \rm \bf
D\rm\underline{59}, 103501 (1999);  \underline{60}, \rm 109903
(1999)(E).

4. O.J.C. Dias and J.P.S. Lemos, hep-th/0410279.

5. R.P. Kerr, \it Phys. Rev. Lett. \rm \rm \underline{11}, 237 (1963).

6. R.C. Myers and M.J. Perry, \it Ann. Phys. \rm \bf
\rm \underline{172}, 304 (1986).

7. S.W. Hawking, C.J. Hunter and M.M. Taylor-Robinson, \it Phys. Rev. \rm \bf
D\rm\underline{59}, 064005 (1999).

8. G.W. Gibbons, H. Lu, D.N. Page and C.N. Pope, \it J. Geom.
Phys. \rm \underline{53}, 49 (2005),
   hep-th/0404008; \it Phys. Rev. Lett. \rm \underline{93}, 171102 (2004), hep-th/0409155.

9. I.S. Booth and R.B. Mann, \it Phys. Rev.  Lett. \rm
\rm \underline{81}, 5052 (1998).

10. D.N. Page, \it Phys. Lett. \rm \bf
B\rm \underline{79}, 235 (1978).

11. J.B. Hartle and S.W. Hawking, \it Phys. Rev. \rm \bf
D\rm\underline{28}, 2960 (1983).

12. G.W. Gibbons, M.J. Perry and C.N. Pope, hep-th/0408217.

13. G.W. Gibbons and S.W. Hawking, \it Phys. Rev. \rm \bf
D\rm\underline{15}, 2752 (1977).

14. M. Cvetic, H. Lu and C.N. Pope, \it Phys. Lett. \rm \bf
B\rm\underline{598}, 273 (2004); hep-th/0407058.

\end{document}